\documentclass[twocolumn,prl,aps,superscriptaddress,showpacs]{revtex4}
\usepackage{graphicx}
\usepackage{bm}

\begin{document}

\title{Tunable terahertz emission from difference-frequency
 in biased superlattices}
\author{Ren-Bao Liu}
\affiliation{Center for Advanced Study, Tsinghua University,
 Beijing 100084, China}
\author{Bang-Fen Zhu}
\affiliation{Center for Advanced Study, Tsinghua University,
 Beijing 100084, China}

\begin{abstract}
The terahertz emission from difference-frequency in biased
superlattices is calculated with the excitonic effect included.
Owing to the doubly resonant condition and the excitonic
enhancement, the typical susceptibility is larger than $10^{-5}$
m/V.  The doubly resonant condition can always be realized by
adjusting the bias voltage and the laser frequencies, thus the
in-situ tunable emission is efficient in the range of 0.5--6
terahertz. Continuous wave operation with $1\%$ quantum
efficiency and $\mu$W output power is feasible while the signal
absorption in undoped superlattices is negligible.
\end{abstract}

\maketitle

Terahertz (THz) electromagnetic waves have potential applications
in many fields, like medical diagnosis, environment monitoring,
high-speed communication, astronomy spectroscopy, etc.
\cite{review}. Though intense tunable THz emission from
free-electron lasers is available in laboratories \cite{FEL},
tunable tabletop THz sources are still desirable for practical
applications. To obtain THz emission from small semiconductor
devices, many schemes have been studied, such as coherent phonons
\cite{phonon}, wave-packet oscillation in asymmetric quantum
wells \cite{QW}, heavy-light hole beatings \cite{beating}, Bloch
oscillations \cite{THzBO}, and difference-frequency in doped
quantum wells \cite{dopedQW}. With state-of-the-art design of
superlattices, a prototype of quantum-cascade THz lasers has been
demonstrated recently \cite{Cascade}. Among these mechanisms for
THz emission, the difference-frequency process is of special
interest because of its in-situ tunability, intense output under
phase-matching condition, and flexibility of operating at both
continuous wave and pulse modes \cite{Kim}. Furthermore, doubly
resonant condition, in which both the input and output are near
resonant with transitions in the system, can also be exploited to
enhance the difference-frequency \cite{dopedQW}.

Doubly resonant difference-frequency in biased superlattices was
also proposed for THz emission \cite{DDDL,free}. Under doubly
resonant condition, the Bloch oscillation is sustained and
amplified by the effective THz potential resulting from the
dipole interaction of excitons and the bichromatic input light,
generating efficient THz radiation \cite{DDDL}. Several
advantages of this method over other difference-frequency schemes
can be expected: First, the applied electric field breaks the
inversion symmetry of the system, leading to a large intraband
dipole matrix element. Secondly, the doubly resonant condition
can always be accomplished by adjusting the static electric field
and tuning the input light. And thirdly, the problem of signal
absorption can also be avoided in undoped superlattices.

Though it has been well-known that the exciton correlation plays
an essential role in THz emission from Bloch oscillation in
optically excited superlattices \cite{QKTHz}, its effect on
difference-frequency in biased superlattices is still unclear.
This question will be addressed in this Letter, and it will be
shown that the excitonic effect can enhance the emission power by
at least two orders of magnitude, which, however, is absent in,
e.g., difference-frequency in doped quantum wells \cite{dopedQW}.

The second-order difference-frequency susceptibility is the key
quantity determining the emission intensity. In principle, it can
be evaluated from the textbook formula derived with the
double-line Feynman diagrams \cite{Shen}. Under the doubly
resonant condition, the difference-frequency susceptibility
\cite{Shen}
\begin{eqnarray}
&&  \chi^{\rm diff}_{jj_1j_2}(\omega=\Omega_1-\Omega_2;
\Omega_1,-\Omega_2)
 \approx  \sum_{a,b}({V\epsilon_0})^{-1}
 \nonumber \\  & & \phantom{\ }
   \times \left[ {(d_{j_2})_{0b}(d_j)_{ba} (d_{j_1})_{a0}
   \over (\varepsilon_{ba}+i\hbar\gamma_1+\hbar\omega)
   (\varepsilon_{a0}-i\hbar\gamma_2-\hbar\Omega_1)}\right.
\nonumber \\  &  & \phantom{\times}
  + \left. {(d_{j_2})_{0b}
   (d_j)_{ba} (d_{j_1})_{a0} \over
    (\varepsilon_{ab}-i\hbar\gamma_1-\hbar\omega)
    (\varepsilon_{b0}+i\hbar\gamma_2-\hbar\Omega_2)}\right],
\label{chidiff}
\end{eqnarray}
where $\Omega_i$ is the frequency of the input light polarized at
${\mathbf e}_{j_i}$ direction, $\varepsilon_{\alpha\alpha'}$
($\alpha, \ \alpha'=a$, $b$, or 0) is the transition energy
between the exciton states $|a\rangle$, $|b\rangle$, and the
semiconductor ground state $|0\rangle$, ${\mathbf
d}_{\alpha\alpha'}$ is the dipole matrix element, $\gamma_2$ and
$\gamma_1$ are the interband and intraband dephasing rates,
respectively, $V$ is the volume of the sample, and $\epsilon_0$
is the vacuum dielectric constant.

With the excitonic effect neglected, the susceptibility of biased
superlattices can be analytically evaluated \cite{free}, and the
result turns out comparable to that of doped quantum wells and
larger by many orders of magnitude than that of bulk
semiconductors \cite{Shen,bulk}. The Coulomb coupling, however,
makes it a formidable task to calculate the susceptibility
directly from Eq. (\protect\ref{chidiff}), since all the
excitonic eigen states should be obtained. To avoid such an
exhausting work, we have developed a time-domain technique, in
which the susceptibility is first transformed into the time
domain, and the result numerically calculated is transformed back
to the frequency domain by standard fast Fourier transformation.

By Fourier transformation of Eq. (\protect\ref{chidiff}), the
time-domain susceptibility of biased superlattices can be derived
as
\begin{eqnarray}
&&\chi^{\rm diff}_{zj_1j_2}(t;t_1,t_2)
 = {1 \over \omega+i\gamma_1}
 {e\over 4\epsilon_0\hbar^3}
\int d{\bm \rho}\sum_{l,m} \nonumber \\ &&\ \  \phantom{-}
m\lambda_m
  \psi_{j_2}^*({\bm \rho},l-m,t-t_2)
  \psi_{j_1}({\bm \rho},l,t-t_1)
\nonumber \\ &&\ \  \times \big[
 \theta(t-t_1) \theta(t_1-t_2)
 e^{-\gamma_1(t-t_1)-\gamma_2(t_1-t_2)}
\nonumber \\ &&\  \ \ + \theta(t-t_2) \theta(t_2-t_1)
 e^{-\gamma_1(t-t_2)-\gamma_2(t_2-t_1)}\big],
\label{chi_t}
\end{eqnarray}
in which the exciton Green's function satisfies the motion
equation in the tight-binding model
\begin{eqnarray}
&& i\hbar\partial_t \psi_j({\bm \rho},l,t)
  = \left[E_g-(2\mu)^{-1}{{\hbar}^2 {\bm \nabla}_{\bm \rho}^2 }\right]
  \psi_j({\bm \rho},l,t)
\nonumber \\ && \ \ \ \ + \sum_{m}\left(leF{D}\delta_{m,0}
 - {\lambda_m /4}\right)\psi_j({\bm \rho},l+m,t)
\nonumber \\ && \ \ \ \
 +V({\bm \rho},l) \psi_j({\bm \rho},l,t)
- \hbar (d_j)_{{\rm cv}}\delta({\bm \rho})\delta_{l,0}\delta(t),
\label{motion}
\end{eqnarray}
where $\lambda_m$ is the tunnelling coefficient between quantum
wells separated by $|m|$ barriers, ${\bm \rho}$ denotes the
in-plane coordinates in real space, $l$ is the index of the unit
cell of the superlattice, $E_g$ is the distance between the
centers of the electron and hole minibands, $\mu$ is the reduced
effective mass of the electron-hole pair, ${D}$ is the
superlattice period, $F$ is the strength of the static field, and
$V({\bm \rho},l)$ is the Coulomb potential. In the derivation of
Eq. (\protect\ref{chi_t}), only the lowest electron and
heavy-hole minibands are included, and the Coulomb potential,
assumed slow-varying as compared to the superlattice potential,
takes the form
\begin{equation}
V({\bm \rho},l)=-e^2(4\pi\epsilon_0\epsilon)^{-1}
\left({\rho^2+l^2{D}^2}\right)^{-1/2},
\end{equation}
where $\epsilon$ is the dielectric constant of the material. It
would not be difficult to include more complexity of realistic
semiconductor systems, such as the valence-band-mixing and the
Coulomb coupling between minibands, which, however, is expected
to modify the results only in details.

The most important feature of Eq. (\protect\ref{chi_t}) is that
the susceptibility has the form of exciton-exciton correlation.
Thus the difference-frequency susceptibility, as well as linear
absorption spectra, can be evaluated by just integrating the
motion equation for the exciton wave-function [Eq.
(\protect\ref{motion})], which can be done with the space-time
difference method proposed by Glutsch {\it et al}. \cite{Glutsch}.
To check the numerical method, the susceptibility has been
numerically calculated with the Coulomb potential artificially
switched off and compared to the analytical result \cite{free},
the deviation is always less than one percent.

\begin{figure}[t]
\begin{center}
\includegraphics[height=7.8cm,width=6.24cm, bb=50 100 570 750,
 clip=true]{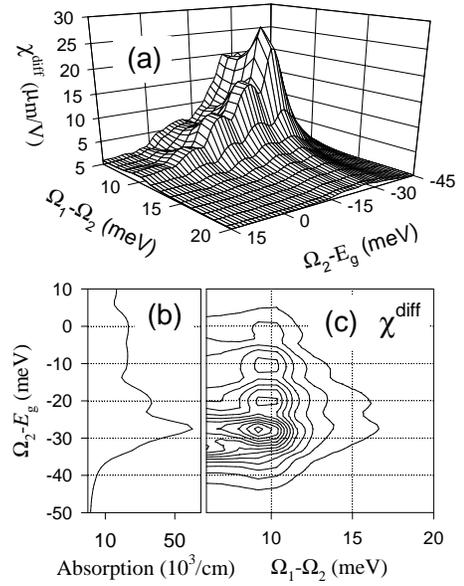}
\end{center}
\caption{(a) The 3D plot of the difference-frequency
susceptibility of a biased superlattice vs. the output frequency
and one of the input frequencies. (b) The linear absorption
spectrum of the superlattice. (c) The contour plot of the
difference-frequency susceptibility.} \label{fig1}
\end{figure}

Now let us focus on the excitonic case. To be specific, here we
consider a superlattice sample that has been well studied for
Bloch oscillation \cite{sample}, namely, a
GaAs/Al$_{0.3}$Ga$_{0.7}$As superlattice with 67 \AA\ well width
and 17 \AA\   barrier width. A Kronig-Penney calculation shows
that the lowest electron and hole minibands have almost perfect
cosinusoid dispersions and the combined miniband width is about 41
meV, so $\lambda_m\approx \delta_{|m|,1}\times 41$ meV. With the
miniband width larger than the emission threshold of optical
phonons, the relaxation in this sample is very rapid, so the
interband and intraband dephasing rates are chosen quite large
values: $\gamma_1^{-1}=0.6$ ps and $\gamma_2^{-1}=0.3$ ps. The
interband dipole matrix element obtained from the ${\mathbf
k}\cdot {\mathbf p}$ theory is $d_{\rm cv}\approx 6.5$ $e$\AA.
Other parameters are such that the excitonic binding energy is
$4.9$ meV, the band gap $E_g=1.511$ eV, the static dielectric
constant $\epsilon=12.9$, and the optical refractive index
$n=3.26$.

\begin{figure}[t]
\begin{center}
\includegraphics[height=5.88cm,width=6.36cm, bb=50 210 580 700,
 clip=true]{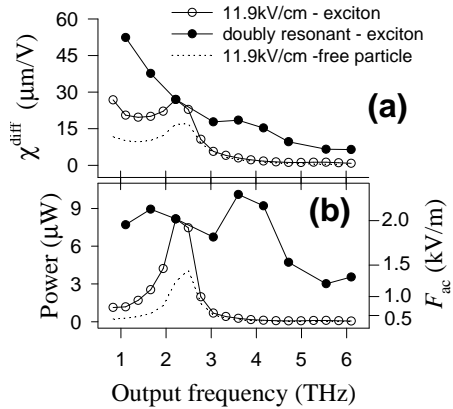}
\end{center}
\caption{ (a) The difference-frequency susceptibility and (b) the
power of the THz emission as functions of the frequency
difference. In (b), the THz field strength $F_{\rm ac}$ is also
indicated. The solid lines with open circles represent excitonic
results at fixed field strength $F=11.9$ kV/cm and optimized
input light frequency, while the solid lines with close circles
correspond to excitonic results with both the static field and
the optical frequencies adjusted to maximize the emission power.
The dotted lines plot the free-particle results at fixed electric
field (11.9 kV/cm) and optimized optical frequencies. For
visibility, the susceptibility and emission power in the
free-particle case are magnified by a factor of 10 and 100,
respectively.} \label{fig2}
\end{figure}

The difference-frequency susceptibility has been calculated for
various static field strength. A typical example is plotted in
Fig. \protect\ref{fig1} for $F=11.9$ kV/cm (correspondingly, the
frequency of the free-particle Bloch oscillation $\nu_{\rm
BO}=2.44$ THz, or $h\nu_{\rm  BO}\equiv eF{D}=10$ meV). The
doubly resonant effect is evident in the peak features of the
susceptibility spectrum. The Coulomb interaction renormalizes the
interband transition energy and the Bloch oscillation frequency.
More importantly, as compared to the free-particle result [see
Fig. \protect\ref{fig2} (a)], the excitonic effect enhances the
susceptibility by more than one order of magnitude. The excitonic
enhancement results mainly from the enhancement of the oscillator
strength at band edge (due to the Sommerfield factor). The
Sommerfield factor also induces extra absorption at band edge,
but considering the fact that the linear optical absorption is
proportional to the second power of the interband dipole matrix
element while the THz signal intensity is proportional to the
fourth power, the net effect of Coulomb interaction is
advantageous to the difference-frequency process.

To estimate the power of the THz emission, we assume perfect
phase-matching for the optical mixing, which can be achieved just
geometrically since the refractive index at THz waveband is larger
than that in optical frequency. The spot size of the laser beams
with appropriate incident angles is taken as $l_x\times l_y$. Thus
the power of the THz signal propagating along the in-plane
$x$-direction is roughly
\begin{equation}
P_{\rm THz}\approx \omega^2 \left|\chi^{\rm diff}\right|^2 N{D}
       \left(8l_y  c^3 \epsilon_0 \epsilon^{1/2}n^2\right)^{-1}
       P_1P_2,
\label{power}
\end{equation}
where $P_1$ and $P_2$ are the power of the two input laser beams,
$N$ is the number of the superlattice periods, and $c$ is the
vacuum light velocity. To be specific, we use following realistic
parameters: $N=50$, $P_1=P_2=0.1$ W, and $l_y=1$ mm.

In Fig. \protect\ref{fig2}, the difference susceptibility and the
THz emission power are plotted against the output frequency both
for a fixed static field and for doubly resonant condition. For
comparison, the results without Coulomb interaction are also
shown for the fixed field strength. The excitonic effect enhances
the emission power by more than two orders of magnitude. Under
doubly resonant condition, the power of THz emission is several
$\mu$W, and the THz electric field strength is larger than 1
kV/m. The efficiency of converting near infrared photons to THz
photons is around $1\%$. Because the doubly resonant condition
can always be realized by simultaneously adjusting the bias
voltage and laser frequencies, the THz emission power is not
sensitive to the frequency difference in the range of 0.5--6 THz,
which, under fixed static field, would otherwise decreases
rapidly as the frequency difference goes away from the Bloch
oscillation frequency.

Negligible absorption of THz signals is another advantage of
using biased superlattices for difference-frequency over other
doubly-resonant schemes, such as that in doped quantum wells. For
parameters in the example above, the optically excited carrier
density, with continuous wave operation mode and $1$ ns$^{-1}$
recombination rate assumed, is of the order of $10^{9}$ cm$^{-2}$
per quantum well, and there is basically no free carriers outside
the laser spot, so the THz signal can propagate without
significant absorption by electrons. Neither is the signal
absorption by optical phonons crucial in the frequency range
considered.

In conclusion, the in-situ tunable THz emission from
difference-frequency in biased superlattices is quite efficient
owing to the tunable doubly-resonant condition, the excitonic
enhancement, and negligible absorption of signals.

\acknowledgments{ The authors are grateful to J. Wu for providing
computer facilities. This work was supported by the National
Science Foundation of China. }

\end{document}